\documentclass[seceq]{ptptex}
\usepackage{wrapft}
\usepackage{graphicx}

\newcommand{\ovl}[1]{\overline{#1}}
\def\beq{\begin{eqnarray}}
\def\eeq{\end{eqnarray}}
\def\bsub{\begin{subequations}}
\def\esub{\end{subequations}}
\def\b{\begin{equation}}

\title{
Effective Potential Approach to Quark Ferromagnetization\\
in High Density Quark Matter
}

\author{
Yasuhiko {\sc Tsue},$^{1,2}$ 
Jo\~ao da {\sc Provid\^encia}$^{3}$,\\
Constan\c{c}a {\sc Provid\^encia}$^{3}$
and Masatoshi {\sc Yamamura}$^{4}$ 
}

\inst{
$^{1}$Center for Computational Physics, Departamento de F\'{i}sica,\\
Universidade de Coimbra, 3004-516 Coimbra, 
Portugal\\
$^{2}$Physics Division, Faculty of Science, Kochi University, Kochi 780-8520, 
Japan\\
$^{3}$Departamento de F\'{i}sica, Universidade de Coimbra, 3004-516 Coimbra, 
Portugal\\
$^4$Department of Pure and Applied Physics, 
Faculty of Engineering Science,\\
Kansai University, Suita 564-8680, Japan
}

\recdate{
\today
}

\abst{
A possibility of spontaneous magnetization in high density symmetric quark matter is investigated 
using the NJL type effective model of 
QCD by means of an effective potential with respect to an auxiliary field.
It is shown that the quark ferromagnetic condensate has non-vanishing value at high baryon density due to 
the tensor-type four-point interaction between quarks. 
}

\begin{document}

\maketitle

\section{Introduction}

One of recent interests about the field governed by the quantum chromodynamics (QCD) may be to 
understand the phase structure on the plane depicted by the temperature and the baryon chemical potential. 
This research area is stimulated by the recent high energy heavy ion collision experiments, in which 
it is reported that the quark-gluon phase is realized.\cite{QGP} 
Further, it is reported in the astrophysical area 
that there exist neutron stars with very large magnetic field which are called magnetars.\cite{magnetar} 
In these extreme conditions at high density and/or at high temperature in the hadronic and/or the quark-gluon matter, 
there may exist various phases such as hadron phase, quark-gluon phase, chiral symmetric or broken phase,\cite{BUBALLA} 
two flavor color-superconducting phase, color-flavor locked phase\cite{cs} and so forth. 
The existence of a ferromagnetic phase has been previously considered by some authors.\cite{Iwazaki,Tatsumi,add1,add2,add3,add4}
In the investigation of the phase structure, especially in the system at finite density, 
various effective models of QCD are used because it is difficult to compute physical quantities at 
finite density directly from QCD, for example, from the lattice QCD simulation method.

The Nambu-Jona-Lasinio (NJL) model\cite{NJL} as an effective model of QCD gives various useful informations 
for the system at finite temperature and density.\cite{HK,Kl,BUBALLA}
Especially, as for the chiral symmetry breaking and restoration, the NJL model has taught various 
properties until now. 
Further, the existence of the color superconducting phase has also been investigated by using the NJL model.\cite{Kitazawa} 
In this paper, with the extension of the original NJL model, we investigate a possibility of the spontaneous spin 
polarization or magnetization
in quark matter at very high density.

Here, let us consider the NJL-type Lagrangian density which has chiral symmetry, namely, the Lagrangian density 
is invariant under the $su(2)_L\times su(2)_R$-chiral transformation.  
Within the four-point interaction between quarks, it is possible for a general form of the Lagrangian density with chiral symmetry to 
include 
scalar, vector and tensor interactions which is written as 
\beq\label{1-1}
& &{\cal L}={\cal L}_{\rm kin}+{\cal L}_{S}+{\cal L}_{V}+{\cal L}_{T}\ , \nonumber\\
& &{\cal L}_{\rm kin}=i{\bar \psi}\gamma^\mu\partial_\mu\psi\ , \nonumber\\
& &{\cal L}_{S}=-G_S\left[
({\bar \psi}\psi)^2+({\bar \psi}i\gamma_5{\vec \tau}\psi)^2\right]\ , \nonumber\\
& &{\cal L}_{V}=-G_V\left[
({\bar \psi}\gamma^{\mu}{\vec \tau}\psi)^2+({\bar \psi}\gamma_5\gamma^{\mu}{\vec \tau}\psi)^2\right]\ , \nonumber\\
& &{\cal L}_{T}=-G_T\left[
({\bar \psi}\gamma^{\mu}\gamma^{\nu}{\vec \tau}\psi)^2+({\bar \psi}i\gamma_5\gamma^{\mu}\gamma^{\nu}\psi)^2\right]\ . 
\eeq
Here, $\psi$ represents the quark field and ${\vec \tau}$ represents the isospin operator. 
The original NJL model Lagrangian density is ${\cal L}_{\rm NJL}={\cal L}_{\rm kin}+{\cal L}_S$.  
Here, ${\bar \psi}\psi$ and ${\bar \psi}i\gamma_5{\vec \tau}\psi$ correspond to the sigma 
meson ($\sigma$) and the pion (${\vec \pi}$) filed, 
respectively, in terms of the bosonized meson fields. 
As is well known, at low energy, the chiral symmetry dynamically breaks down and the chiral condensate which is 
the vacuum expectation value of ${\bar \psi}\psi$ becomes non-zero.
Similarly to the original NJL model, 
the field ${\bar \psi}\gamma^{\mu}{\vec \tau}\psi$ and ${\bar \psi}\gamma_5\gamma^{\mu}{\vec \tau}\psi$ in ${\cal L}_V$  
corresponds to the $\rho$ meson and $a$ meson fields, respectively, if these composite fields are treated by means of a bosonization. 
However, it is known that the physical $a$ meson field is constructed from the original $a$ meson field 
by mixing with ${\pi}$ and ${\rho}$ mesons due to the chiral symmetry breaking with the chiral condensate $\langle {\bar \psi}\psi\rangle$, 
which results to the mass difference between $\rho$ and $a$ mesons.\cite{meissner} 
In addition to the above four-point interactions, further, 
from the structure of the Dirac gamma matrices, the tensor interaction part, ${\cal L}_T$, should be introduced. 
Here, ${\bar \psi}\gamma^{\mu}\gamma^{\nu}{\vec \tau}\psi$ may correspond to the $a_2$ meson field. 
In the nonrelativistic limit, this tensor meson exchange interaction by $a_2$ meson leads to the tensor force, the spin-orbit force, 
the second spin-orbit force and so on.\cite{Tsushima}
This term may also be understood in the framework of a Fierz transformation of a standard NJL interaction.

In this paper, we consider a system at high density such as the interior of the neutron star, in which 
there may exist quark matter at high density, and/or the quark star\cite{Qstar}. 
At high baryon density, the chiral symmetry is restored and the dynamical quark mass is zero, namely, 
the chiral condensate $\langle{\bar \psi}\psi\rangle$ is equal to zero. 
In this case, the non-vanishing vacuum expectation values of ${\bar \psi}\psi$, ${\bar \psi}i\gamma_5{\vec \tau}\psi$, 
${\bar \psi}\gamma^{\mu}{\vec \tau}\psi$ and ${\bar \psi}\gamma_5\gamma^{\mu}{\vec \tau}\psi$ do not appear, 
that is, the condensates are zero for these composite field operators. 
Especially, as for the spin polarization originated from the axial vector interaction, 
it may be possible to realize the spin polarization $\langle {\bar \psi}\gamma_5\gamma^3{\vec \tau}\psi\rangle$.\cite{add2,add4}
However, at high baryon density, where the chiral symmetry is restored and the quark mass is zero, the spin 
polarization does not appear.\cite{add2}
Therefore, it is regarded that these terms only cause the excitation modes around the vacuum which correspond to meson excitation modes of  
$\sigma$, ${\pi}$, ${\rho}$ and $a$ mesons in the low energy vacuum with a finite chiral condensate.
Thus, we could safely discard the ${\cal L}_S$ and ${\cal L}_V$ when we only investigate the vacuum at high density 

In Refs.\citen{Bohr1} and \citen{Bohr2}, by two of the present authors (J. da P. and C. P.), 
a possibility of the ferromagnetic condensate in the quark matter 
at high density has already been considered by means of the thermodynamical treatment of the NJL model 
with tensor interaction. 
In this paper, we reinvestigate the ferromagnetic condensate by a 
field theoretical approach with the aim of the extension to the finite temperature system. 
Namely, we derive the effective potential for the ferromagnetic condensate.

This paper is organized as follows. 
In the next section, the effective potential for the quark ferromagnetic condensate 
is derived by using the auxiliary field method. 
In \S 3, the analytical form of the effective potential is given at zero temperature. 
In \S 4, the numerical results are given and it is shown that the quark ferromagnetic condensate is really realized for 
high density quark matter. 
Further, the implication to the magnetars is discussed. 
The last section is devoted to a summary. 
In Appendix A, the effective potential is reinvestigated in terms of another normalization. 
In Appendix B, the effect of the vacuum polarization is summarized briefly.

\section{Effective potential approach to symmetric quark matter with tensor-type interaction}

In this section, we derive an effective potential with respect to the quark ferromagnetic condensate. 
As is mentioned in \S 1, since we pay attention to the vacuum at high density quark matter, 
the chiral condensate and the excited modes may be discarded safely. 
Thus, let us start with the following Lagrangian density: 
\beq\label{2-1}
{\cal L}=i{\bar \psi}\gamma^\mu\partial_\mu\psi-\frac{G}{4}({\bar \psi}\gamma^{\mu}\gamma^{\nu}{\vec \tau}\psi)
({\bar \psi}\gamma_{\mu}\gamma_{\nu}{\vec \tau}\psi)\ , 
\eeq
where we omit the second term of ${\cal L}_T$ and the coupling constant $G_T$ is redefined as $G_T=G/4$. 
In the Dirac representation for the Dirac gamma matrices, we get 
\beq\label{2-2}
\gamma^1\gamma^2=-i\Sigma_3=
-i\left(
\begin{array}{cc}
\sigma_3 &  0 \\
0 & \sigma_3 \\
\end{array}
\right) \ . 
\eeq
Here, $\sigma_3$ is the third component of the Pauli matrix. 
If the vacuum expectation value of ${\bar \psi}\gamma^1\gamma^2{\vec \tau}\psi$ has non-zero value, 
the spin of quarks is aligned along the third axis which leads to the magnetization due to the quark magnetic moment.
Thus, it is enough to investigate a possibility of spin alignment along the third axis. 
The generating functional of the Green function is expressed in the form of 
the Feynman path integral as
\beq\label{2-3}
Z\propto \int{\cal D}{\bar \psi}{\cal D}\psi
\exp\left[i\int d^4x\left(
{\bar \psi}i\gamma^\mu\partial_\mu\psi+\frac{G}{2}
({\bar \psi}\Sigma_3\tau_k\psi)
({\bar \psi}\Sigma_3\tau_k\psi)\right)\right]\ . 
\eeq
Following the standard technique of the auxiliary field method, we introduce the auxiliary field $F_k$, 
which corresponds to the ferromagnetic condensate. 
First, the functional Gaussian integration, 
\beq\label{2-4}
1=\int{\cal D}F_k\exp\left[
-\frac{i}{2}\int d^4 x\left(
F_k+G({\bar \psi}\Sigma_3\tau_k\psi)\right)
G^{-1}\left(F_k+G({\bar \psi}\Sigma_3\tau_k\psi)\right)\right]\ , 
\eeq
is inserted in the expression of the generating functional in Eq.(\ref{2-3}). 
Apart from the overall normalization factor, we obtain
\beq\label{2-5}
Z&\propto&\int{\cal D}{\bar \psi}{\cal D}\psi{\cal D}F_k
\exp\left[i\int d^4x\left(
{\bar \psi}i\gamma^{\mu}\partial_\mu\psi-\frac{1}{2G}F_k^2
-F_k({\bar \psi}\Sigma_3\tau_k\psi)\right)\right]
\nonumber\\
&=&\int{\cal D}F_k\exp\left[i\int d^4x
\left(-\frac{1}{2G}F_k^2\right)+\ln {\rm Det}[i\gamma^{\mu}\partial_{\mu}-F_k\Sigma_3\tau_k]\right]
\nonumber\\
&=&\int{\cal D}F_k\exp\Biggl[i\int d^4x
\left(-\frac{1}{2G}F_k^2\right)
\nonumber\\
& &\qquad\qquad\qquad
+\frac{1}{2}{\rm Tr}\left[\ln \left(\gamma^\mu p_\mu-F_k\Sigma_3\tau_k\right)+
\ln\left(-\gamma^\mu p_\mu-F_k\Sigma_3\tau_k\right)\right]\Biggl]
\nonumber\\
&=&\int{\cal D}F_k\exp\Biggl[i\int d^4x
\left(-\frac{1}{2G}F_k^2\right)
\nonumber\\
& &\qquad\qquad\qquad
+\frac{1}{2}{\rm Tr}\ln \left[-p^2+(F_k\tau_k)^2-F_k p_\mu\gamma^\mu\Sigma_3\tau_k
+F_k\Sigma_3p_\mu\gamma^\mu\tau_k\right]
\Biggl] . \ \ \ 
\eeq
Here, from the first line to the second line, we carried out the 
fermionic path integral and from the second line to the third line, 
the term in which the momentum $p_{\mu}$ reverses to $-p_{\mu}$ is added, which gives the same contribution of the term 
with $p_{\mu}$. 
Then, we divide them by factor 2.
Further, from the third line to the fourth line, we used the relation
$\gamma^\mu p_\mu\gamma^\nu p_\nu=p^2$. 
From the first line, the equation for $F_k$ gives $F_k=-G\langle {\bar \psi}\Sigma_3\tau_k
\psi\rangle$. 
Thus, the vacuum expectation value of $F_k$, that is the mean filed, represents the 
quark spin alignment or quark ferromagnetic condensate.

The last two terms in the logarithmic function on the last line in Eq.(\ref{2-5}) except for $F_k\tau_k$ 
is recast into 
\beq\label{2-6}
-p_\mu\gamma^\mu\Sigma_3+\Sigma_3p_\mu\gamma^\mu
=-2i(p_1\gamma^2-p_2\gamma^1)\ ,
\eeq
where we used the relation $[\sigma_i , \sigma_j]=i2\sigma_k$ for 
the cyclic $(i, j, k)=(1, 2, 3)$. 
Thus, the generating functional $Z$ is rewritten as 
\beq\label{2-7}
Z\propto
\int{\cal D}F_k\exp\left[
i\int d^4x\left(
-\frac{F_k^2}{2G}+\frac{1}{2i}
{\rm tr}\ln\left[
-p^2+(F_k\tau_k)^2-2i(p_1\gamma^2-p_2\gamma^1)F_k\tau_k\right]\right)\right]
\ . \nonumber\\
& &
\eeq
Next, let us calculate the logarithmic term. 
We can calculate it as follows: 
\beq\label{2-8}
& &{\rm tr}\ln\left[
-p^2+(F_k\tau_k)^2-2i(p_1\gamma^2-p_2\gamma^1)F_k\tau_k\right]
\nonumber\\
&=&\frac{1}{2}{\rm tr}\ln(-p^2+(F_k\tau_k)^2)^2
+{\rm tr}\ln\left(\!
1-\frac{2i}{-p^2+(F_k\tau_k)^2}(p_1\gamma^2-p_2\gamma^1)F_k\tau_k\!\right) . \ \ 
\eeq
Here, by using the expansion of the logarithmic function 
$\ln(1-x)=-x-\frac{1}{2}x^2-\frac{1}{3}x^3-\cdots$, 
the second term in Eq.(\ref{2-8}) is recast into 
\beq\label{2-9}
& &
{\rm tr}\ln\left(
1-\frac{2i}{-p^2+(F_k\tau_k)^2}(p_1\gamma^2-p_2\gamma^1)F_k\tau_k\right)
\nonumber\\
&=&{\rm tr}
\biggl[
-\frac{1}{2}\left(\frac{(2iF_k\tau_k)^2}{(-p^2+(F_k\tau_k)^2)^2}\right)(-(p_1^2+p_2^2))
\nonumber\\
& &\ \ \ \ 
-
\frac{1}{4}\left(\frac{(2iF_k\tau_k)^2}{(-p^2+(F_k\tau_k)^2)^2}\right)^2
(-(p_1^2+p_2^2))^2-\cdots\biggl]
\nonumber\\
&=&
\frac{1}{2}{\rm tr}
\ln\left[1-\left(\frac{2F_k\tau_k\sqrt{p_1^2+p_2^2}}{-p^2+(F_k\tau_k)^2}
\right)^2\right]\ , 
\eeq
where we used the relation 
\beq\label{2-10}
& &(p_1\gamma^2-p_2\gamma^2)^2
=-(p_1^2+p_2^2)\ , \nonumber\\
& &{\rm tr}(p_1\gamma^2-p_2\gamma^1)=0
\eeq
with $(\gamma^1)^2=(\gamma^2)^2=-1$. 
Thus, finally, we get the generating functional $Z$ from Eq.(\ref{2-7}) with (\ref{2-8}) 
and (\ref{2-9}) as 
\beq\label{2-11}
Z&\propto&
\int{\cal D}F_k\exp\Biggl[
i\int d^4 x\biggl(-\frac{F_k^2}{2G}
+\frac{1}{4i}
{\rm tr}\ln\left(-p^2+(F_k\tau_k)^2-2F_k\tau_k\sqrt{p_1^2+p_2^2}\right)
\nonumber\\
& &\qquad\qquad\qquad\qquad
+\frac{1}{4i}
{\rm tr}\ln\left(-p^2+(F_k\tau_k)^2+2F_k\tau_k\sqrt{p_1^2+p_2^2}\right)\!\!
\biggl)\Biggl]\ . 
\eeq

It should be here noted that the argument of the logarithmic function in Eq.(\ref{2-11}) 
can be expressed as 
\beq\label{2-12}
& &-p^2+(F_k\tau_k)^2\pm2F_k\tau_k\sqrt{p_1^2+p_2^2}
=-p_0^2+\epsilon_p^{(\pm)2}\ , \nonumber\\
& &
\epsilon_p^{(\pm)}\equiv
\left({\mib p}^2+(F_k\tau_k)^2\pm 2F_k\tau_k\sqrt{p_1^2+p_2^2}
\right)^{\frac{1}{2}}
=\sqrt{\left((F_k\tau_k)\pm\sqrt{p_1^2+p_2^2}\right)^2+p_3^2}\ . \nonumber\\
& &
\eeq
Here, $\epsilon_p^{(\pm)}$ is identical with the energy eigenvalue of the 
Dirac equation for $\psi$ derived from the first line of Eq.(\ref{2-5}) shown in Ref.\citen{Bohr1}.

Since the expectation values of $\langle \tau_1\rangle$ and $\langle \tau_2\rangle$ for $u$ and $d$ 
quark field are zero, we retain only $F_3$ which we denote as $F$ simply. 
Thus, hereafter, we adopt $k=3$ and we denote $F_k\tau_k$ as $F\epsilon_\tau$ where 
$\epsilon_{\tau}$ is the eigenvalue of $\tau_3$ for $u$ and $d$ quark field which gives 
$\epsilon_\tau=\pm 1$ as was shown in Ref.\citen{Bohr1}. 
In general, the effective action $\Gamma[F]$ is defined and introduced as 
\beq\label{2-13}
Z=\exp\left(i\Gamma[F]\right)\ . 
\eeq
Further, the effective potential $V[F]$ is defined and introduced here as 
\beq\label{2-14}
V[F]=-\frac{\Gamma[F]}{\int d^4 x}\ . 
\eeq
In our case, the effective potential $V[F]$ is already obtained in 
Eq.(\ref{2-11}) in the above calculation:
\beq\label{2-15}
V[F]&=&
\frac{F^2}{2G}-
N_c\int\frac{d^4 p}{i(2\pi)^4}
\biggl[\ln\left(
-p_0^2+{\mib p}^2+(F\epsilon_\tau)^2-2F\epsilon_\tau\sqrt{p_1^2+p_2^2}\right)
\nonumber\\
& &\qquad\qquad\qquad\qquad
+\ln\left(
-p_0^2+{\mib p}^2+(F\epsilon_\tau)^2+2F\epsilon_\tau\sqrt{p_1^2+p_2^2}\right)
\biggl] . \ \ 
\eeq
Here, trace is taken with respect to 
color and Dirac indices, which leads to 
$N_c\times 4$. 
As for the flavor, we fix $\epsilon_{\tau}=1$ and multiply by 2 
because the same contribution is obtained for $\epsilon_{\tau}=1$ and $-1$. 
Further, the part of integration is rewritten as follows:  
\beq\label{2-16}
V[F]
&=&\int^F\!\! \frac{\delta V[F]}{\delta F}dF \nonumber\\
&=&
\frac{F^2}{2G}-
4N_c\int^F\!\! dF\int\!\!\frac{d^4 p}{i(2\pi)^4}
\left[\frac{F-\sqrt{p_1^2+p_2^2}}{-p_0^2+\epsilon_{p}^{(-)2}}
+\frac{F+\sqrt{p_1^2+p_2^2}}{-p_0^2+\epsilon_{p}^{(+)2}}\right]
\ .\nonumber\\
& &
\eeq

In order to extend the above treatment to a finite density case, 
the quark chemical potential $\mu$ should be introduced in the 
starting Lagrangian density: 
\beq\label{2-17}
{\cal L}\longrightarrow {\cal L}+\mu\psi^{\dagger}\psi
={\cal L}+{\bar \psi}\mu\gamma^0\psi\ . 
\eeq
Therefore, we replace $p_0$ as $p_0+\mu$. 
Further, to extend to the finite temperature system, 
the Matsubara formalism may be used.\cite{Matsubara} 
We replace $p_0$ and $p_0$-integration into the following: 
\beq\label{2-18}
\int\frac{d^4 p}{i(2\pi)^4}f(p_0,{\mib p})
\longrightarrow 
T\sum_{n=-\infty}^{\infty}\int \frac{d^3 {\mib p}}{(2\pi)^3}
f(i\omega_n+\mu,{\mib p})\ , 
\eeq
where $\omega_n\equiv (2n+1)\pi T$ is the Matsubara frequency and $n$ is an integer. 
Here, $T$ represents temperature and $\mu$ is a quark chemical potential.

In the finite temperature and chemical potential system, 
we use the replacement (\ref{2-18}) for the above effective potential in Eq.(\ref{2-16}). 
Here, the Matsubara sum is taken by the standard method\cite{KB}, which results
\beq\label{2-19}
T\sum_{n=-\infty}^{\infty}\int\frac{d^3{\mib p}}{(2\pi)^3}
\frac{F\pm\sqrt{p_1^2+p_2^2}}
{(\omega_n-i\mu)^2+\epsilon_p^{(\pm)2}}
&=&
\frac{1}{2}\int\frac{d^3 {\mib p}}{(2\pi)^3}
\frac{F\pm\sqrt{p_1^2+p_2^2}}
{2\epsilon_p^{(\pm)2}}
\left(n_-^{(\pm)}-n_+^{(\pm)}\right)\ . \nonumber\\
& &
\eeq
As a result, the effective potential at finite temperature and 
chemical potential is derived as 
\beq\label{2-20}
V[F]&=&\frac{F^2}{2G}
+2N_c\int^F\!\! dF
\int\!\!\frac{d^3 {\mib p}}{(2\pi)^3}
\biggl[
\frac{F-\sqrt{p_1^2+p_2^2}}
{\sqrt{{\mib p}^2+F^2-2F\sqrt{p_1^2+p_2^2}}}
\left(n_+^{(-)}-n_-^{(-)}\right)
\nonumber\\
& &\qquad\qquad\qquad\quad
+\frac{F+\sqrt{p_1^2+p_2^2}}
{\sqrt{{\mib p}^2+F^2+2F\sqrt{p_1^2+p_2^2}}}
\left(n_+^{(+)}-n_-^{(+)}\right)
\biggl]\ , 
\eeq
where $n_{\pm}^{(\pm)}$ is defined by  
\beq\label{2-21}
n_{\pm}^{(\pm)}
=\frac{1}{\exp\left(\frac{\pm\epsilon_p^{(\pm)}-\mu}{T}\right)+1}\ .
\eeq
Of course, $n_{\pm}^{(\pm)}$ is identical to the Fermi distribution function. 
Here, $\epsilon_p^{(\pm)}$ has been already defined 
in Eq.(\ref{2-12}) and it also appears in the denominator of the above integrand.

Finally, let us give the ``gap" equation which is derived by $\delta V[F]/\delta F=0$, that is, 
\beq\label{2-22}
\frac{\delta V[F]}{\delta F}
&=&\frac{F}{G}+
2N_c\int\!\!\frac{d^3{\mib p}}{(2\pi)^3}
\biggl[
\frac{F-\sqrt{p_1^2+p_2^2}}
{\sqrt{{\mib p}^2+F^2-2F\sqrt{p_1^2+p_2^2}}}
\left(n_+^{(-)}-n_-^{(-)}\right)
\nonumber\\
& &\qquad\qquad\qquad\qquad\qquad
+\frac{F+\sqrt{p_1^2+p_2^2}}
{\sqrt{{\mib p}^2+F^2+2F\sqrt{p_1^2+p_2^2}}}
\left(n_+^{(+)}-n_-^{(+)}\right)
\biggl] \nonumber\\
&=&0\ . 
\eeq
From the above gap equation, the ferromagnetic condensate $F$ is evaluated.

\section{Symmetric quark matter at finite density and zero temperature}

In this paper, we consider symmetric quark matter at finite density and zero temperature. 
It should be noted that $n_-^{(\pm)}$ represents the negative energy contribution. 
Here, we should replace $n_-^{(\pm)}$ into $n_-^{(\pm)}-1$ if the negative energy contribution is omitted. 
It is possible to introduce the three-momentum cutoff $\Lambda$ to take into account of the negative energy contribution. 
The result for the phase transition point and so on is only obtained by replacing $1/G$ into a renormalized coupling 
$1/G_r=1/G-\Lambda^2/\pi^2$. 
Therefore, the qualitative behavior is not changed in the later discussion, while the vacuum polarization is important 
for the spin polarization due to other interactions,\cite{add1} especially axial vector interaction.
This procedure is described in Appendix B.
Under this replacement, the relation 
$n_-^{(\pm)}-1=0$ is obtained at zero temperature.
Thus, at zero temperature, we can express 
$n_+^{(\pm)}=\theta(\mu-\epsilon_p^{(\pm)})$ where 
$\theta(x)$ is the Heaviside step function or the theta function. 
In this case, the effective potential (\ref{2-20}) is written as 
\beq\label{3-1}
V[F]&=&\frac{F^2}{2G}
+{2N_c}\int^F\!\! dF
\int\!\!\frac{d^3 {\mib p}}{(2\pi)^3}
\biggl[
\frac{F-\sqrt{p_1^2+p_2^2}}
{\sqrt{{\mib p}^2+F^2-2F\sqrt{p_1^2+p_2^2}}}
\theta(\mu-\epsilon_p^{(-)})
\nonumber\\
& &\qquad\qquad\qquad\qquad\qquad\quad
+\frac{F+\sqrt{p_1^2+p_2^2}}
{\sqrt{{\mib p}^2+F^2+2F\sqrt{p_1^2+p_2^2}}}
\theta(\mu-\epsilon_p^{(+)})
\biggl] \ .\nonumber\\
\eeq
Without loss of generality, $F$ is assumed to be positive since the effective potential is an even function with respect to $F$. 
The role of the theta function in Eq.(\ref{3-1}) is to give the Fermi surface, in which 
the condition for the allowed momentum region is presented. 
For example, $p_3$ should satisfy the relation
\beq\label{3-2}
-\mu \leq p_3 \leq \mu\ .
\eeq
Also, the magnitude of the remaining momentum components, $\sqrt{p_1^2+p_2^2}$, should be satisfied:
\beq\label{3-3}
& &-\mu \leq F-\sqrt{p_1^2+p_2^2} \leq \mu\ , \quad
{\rm i.e.,}\quad
F-\mu \leq \sqrt{p_1^2+p_2^2}\leq F+\mu\qquad ({\rm for}\ \ \epsilon_p^{(-)})\ , \nonumber\\
& &-\mu \leq F+ \sqrt{p_1^2+p_2^2} \leq \mu\ , \quad
{\rm i.e.,}\quad
0 \leq \sqrt{p_1^2+p_2^2}\leq \mu-F \qquad ({\rm for}\ \ \epsilon_p^{(+)})\ .  
\eeq 
Here, it is understood that the Fermi surface is modified from spherical form. 
Therefore, the effective potential should be calculated in the case 
$F<\mu$ and $F>\mu$ separately.

\subsection{$F>\mu$}

In the case $F>\mu$, from the lower relation in Eq.(\ref{3-3}), 
$\sqrt{p_1^2+p_2^2}$ is not allowed to take a finite value except for 0. 
Thus, for $\epsilon_p^{(+)}$, the Fermi surface is closed. 
As a result, the second term of the integrand in Eq.(\ref{3-1}) does not give any contribution to 
the effective potential. 
Thus, the effective potential is computed as 
\beq\label{3-4}
V[F]&=&\frac{F^2}{2G}+2N_c\int_{\mu}^F dF\int\frac{d^3{\mib p}}{(2\pi)^3}
\left[\frac{F-\sqrt{p_1^2+p_2^2}}{\sqrt{{\mib p}^2+F^2-2F\sqrt{p_1^2+p_2^2}}}\theta(\mu-\epsilon_p^{(-)})\right]
\nonumber\\
&=&\frac{F^2}{2G}-\frac{N_c}{6\pi}\mu^3 F +\frac{N_c\mu^4}{6\pi}\ . 
\eeq
The quark number density $\rho_q$ is similarly calculated as 
\beq\label{3-5}
\rho_q&=&N_cN_f\int\frac{d^3{\mib p}}{(2\pi)^3}\theta(\mu-\epsilon_p^{(-)})
=\frac{N_c}{2\pi}F\mu^2\ 
\eeq
with $N_f=2$.

From the gap equation $\delta V/\delta F=0$, if there exists a potential minimum in the region with $F>\mu$, then 
the solution of the gap equation, $F_{\rm min}$, is obtained as 
\beq\label{3-6}
F_{\rm min}=\frac{N_cG\mu^3}{6\pi}\ . 
\eeq 
Then, the quark number density which gives the minimum of the effective potential is expressed in terms of 
the quark chemical potential $\mu$ as 
\beq\label{3-7}
\rho_q=\frac{N_c^2G}{12\pi^2}\mu^5\ . 
\eeq

\subsection{$F<\mu$}

In the case $F<\mu$, the Fermi surface is open for $\epsilon_p^{(\pm)}$. 
Thus, the second term of integrand in Eq.(\ref{3-1}) gives a contribution to the effective potential as well as 
the first term considered in the case $F>\mu$.  
Taking care of the range of integration and writing $\sqrt{p_1^2+p_2^2}=p_{\perp}$, we obtain  
\beq\label{3-8}
V[F]&=&\frac{F^2}{2G}+2N_c\int_0^F dF\Biggl[
\frac{1}{2\pi^2}\int_0^{\mu+F} dp_{\perp} p_{\perp}(F-p_{\perp})\ln\left(\frac{\mu+\sqrt{\mu^2-(F-p_{\perp})^2}}{F-p_{\perp}}\right)
\nonumber\\
& &\qquad\qquad\qquad\qquad
+\frac{1}{2\pi^2}\int_0^{\mu-F} dp_{\perp} p_{\perp}(F+p_{\perp})\ln\left(\frac{\mu+\sqrt{\mu^2-(F+p_{\perp})^2}}{F+p_{\perp}}\right)
\Biggl]\nonumber\\
&=&\frac{F^2}{2G}-
\frac{N_c}{3\pi^2}\Biggl[
\frac{\sqrt{\mu^2-F^2}}{4}(3F^2\mu+2\mu^3)+F\mu^3{\rm arctan}\frac{F}{\sqrt{\mu^2-F^2}}\nonumber\\
& &\qquad\qquad\qquad
-\frac{F^4}{4}
\ln\frac{\mu+\sqrt{\mu^2-F^2}}{F}-\frac{\mu^4}{2}\Biggl]\ .
\eeq
It should be here noted that the above effective potential is normalized so as to be 
equal to zero when $F=0$, namely $V[F=0]=0$, due to the lower limit of integration being 0. 
The effective potential derived here with the above-mentioned normalization corresponds to calculating 
$V[F]-V[F=0]$. 
If it is necessary to get $V[F]$ itself, we have to add $V[F=0]$ which implies the contribution of the 
free massless quark gas and diverges due to the vacuum fluctuation.\cite{HTF} 
In Appendix A, $V[F]$ itself is given so as to avoid a divergence.

The quark number density is derived as 
\beq\label{3-9}
\rho_q&=&
N_cN_f\int\frac{d^3{\mib p}}{(2\pi)^3}\left(\theta(\mu-\epsilon_p^{(-)})+\theta(\mu-\epsilon_p^{(+)})\right)\nonumber\\
&=&\frac{N_cN_f}{2\pi^2}\int_0^{\mu+F}\!\!dp_{\perp}p_{\perp}\sqrt{\mu^2-(F-p_{\perp})^2}
+\frac{N_cN_f}{2\pi^2}\int_0^{\mu-F}\!\!dp_{\perp}p_{\perp}\sqrt{\mu^2-(F+p_{\perp})^2}\nonumber\\
&=&\frac{N_cN_f}{6\pi^2}
\left[\sqrt{\mu^2-F^2}(F^2+2\mu^2)+3F\mu^2{\rm arctan}\frac{F}{\sqrt{\mu^2-F^2}}\right]
\eeq
The gap equation $\delta V/\delta F=0$ has always a trivial solution, $F=0$. 
If there exists the potential minimum in $F<\mu$ except for $F=0$, then the gap equation is rewritten as
\beq\label{3-10}
& &F_{\rm min}=\left\{
\begin{array}{l}
0 \\
{\displaystyle 
\frac{N_cG}{3\pi^2}}\Biggl[
2F_{\rm min}\mu\sqrt{\mu^2-F_{\rm min}^2}+\mu^3{\rm arctan}\frac{F_{\rm min}}{\sqrt{\mu^2-F_{\rm min}^2}}
-F_{\rm min}^3\ln\frac{\mu+\sqrt{\mu^2-F_{\rm min}^2}}{F_{\rm min}}\Biggl] \ .
\end{array}\right. \nonumber\\
& &
\eeq
Then, the quark number density which gives the minimum of the effective potential is expressed in terms of 
the quark chemical potential $\mu$ as
\beq\label{3-11}
& &\rho_q=\left\{
\begin{array}{ll}
\displaystyle \frac{2N_c\mu^3}{3\pi^2} &\!\!\!\! {\rm for}\ F_{\rm min}=0 \\
{\displaystyle 
\frac{N_c}{3\pi^2}}\Biggl[
(F_{\rm min}^2+2\mu^2)\sqrt{\mu^2-F_{\rm min}^2}+3F_{\rm min}\mu^2{\rm arctan}\frac{F_{\rm min}}{\sqrt{\mu^2-F_{\rm min}^2}}
\Biggl] &\!\!\!\! {\rm for}\ F=F_{\rm min} \ . \\
\end{array}\right. \nonumber\\
& &
\eeq

\section{Numerical result}

\begin{figure}[b]
\begin{center}
\includegraphics[height=5.4cm]{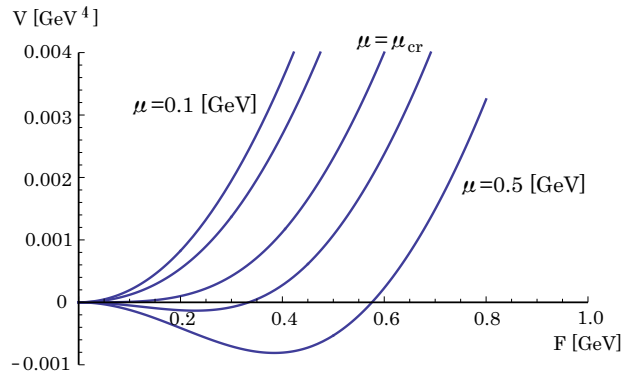}
\caption{
The effective potential is shown as a function of $F$.
From top, $\mu=0.1$ GeV, 0.3 GeV, $\mu_{\rm cr}$, 0.45 GeV and 0.5 GeV, respectively.
}
\label{fig:4-1}
\end{center}
\end{figure}

As is well known, the NJL model is a nonrenormalizable model. 
Thus, in the original NJL model, the momentum cutoff parameter $\Lambda$ and the strength of the scalar and pseudoscalar 
interactions $G_S$ are necessary and is taken so as to reproduce the quark mass and pion decay constant. 
Usually, three-momentum cutoff $\Lambda=0.631$ GeV and $G_S=0.214$ fm${}^2=5.514$ GeV${}^{-2}$ are adopted.\cite{HK}
In our calculation, it is not necessary to introduce the momentum cutoff explicitly because 
the quark chemical potential $\mu$, which corresponds to the Fermi energy at zero temperature, 
plays a role of momentum cutoff. 
Thus, in this model, only one parameter is included, namely $G$, which is the strength of the 
tensor interaction.

As for the parameter $G$, if the Fierz transformation\cite{CL} from the original NJL model is assumed, then 
the following relation is obtained:
\beq\label{4-1}
G_S\left[({\bar \psi}\psi)^2+({\bar \psi}i\gamma_5{\vec \tau}\psi)^2\right]
=\frac{G_S}{4}\left[({\bar \psi}\psi)^2-\frac{1}{2}({\bar \psi}\gamma^{\mu}\gamma^{\nu}{\vec \tau}\psi)^2
+\cdots\right] \ .
\eeq
Then, in this case, $|G_S/G_T|=2$ is obtained. 
However, in this paper, we regard $G$ as a free parameter. 
As a guide to fix the parameter $G$, it is noted that $G_T\equiv G/4$ should be compared with 
$G_S$. 
Thus, we take $G_T=5$ GeV${}^{-2}$, namely $G=20$ GeV${}^{-2}$, which is comparable with the original NJL model parameter $G_S$.

In Fig.\ref{fig:4-1}, the effective potential in Eqs.(\ref{3-8}) and (\ref{3-4}) for $F<\mu$ and $F>\mu$, respectively, 
is shown as a function of $F$ with $G=20$ GeV${}^{-2}$. 
From top, $\mu=0.1$ GeV, 0.3 GeV, $\mu_{\rm cr}$, 0.45 GeV and 0.5 GeV, respectively, are shown. 
In $\mu>\mu_{\rm cr}=0.406$ GeV, the ferromagnetic condensate $F$ exists with non-vanishing value. 
Thus, the quark spin is aligned and the magnetic field appears spontaneously. 
The critical baryon density, which corresponds to one third of the critical quark number density, is 
about 3.47 times the normal nuclear density.  
Here, the critical quark chemical potential $\mu_{\rm cr}$ is obtained by $\delta^2 V[F=0]/\delta F^2=0$. 
Thus, the critical quark number density or baryon density is calculated by using the relation between 
the quark chemical potential and quark number density given in Eq.(\ref{3-9}) or (\ref{3-5}). 
The critical density is varied as the parameter $G$ is changed. 
In Table \ref{table:1}, the critical baryon densities are collected.  

\begin{table}[b]
\caption{The critical baryon density $\rho_{\rm cr}$ and critical quark chemical potential $\mu_{\rm cr}$.}
\label{table:1}
\begin{center}
\begin{tabular}{c|c|c} \hline
$G$ / GeV${}^{-2}$ & $\rho_{\rm cr}/\rho_0$ & $\mu_{\rm cr}$ / GeV\\ \hline\hline
15 & 5.34 & 0.468 \\
20 & 3.47 & 0.406 \\
25 & 2.48 & 0.363\\
\hline
\end{tabular}
\end{center}
\end{table}

\begin{figure}[t]
\begin{center}
\includegraphics[height=5.4cm]{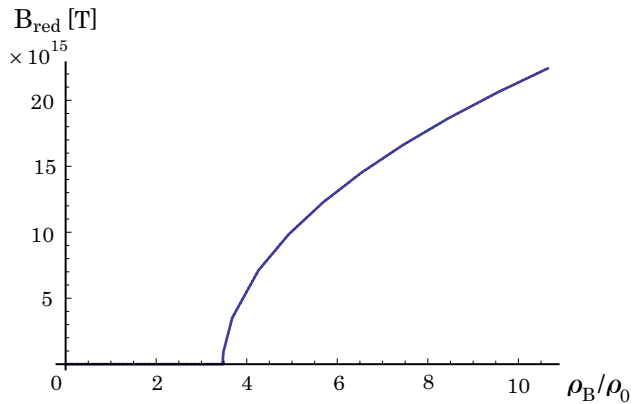}
\caption{
The strength of ``reduced" magnetic field is shown as a function of $\rho_B/\rho_0$.
}
\label{fig:4-2}
\end{center}
\end{figure}

As is mentioned in \S 2, non-vanishing $F$ means that there is a spin alignment of quarks, 
which leads to the magnetic field due to the quark magnetic moment. 
Thus, it means that there is a possibility that spontaneous magnetization occurs in the quark matter at high baryon density.
We roughly estimate the strength of the magnetic field. 
The condensate $F$ has the dimension of energy. 
Here, we introduce and define a ``reduced" magnetic field $B_{\rm red}$ as 
\beq\label{4-2}
{\ovl \mu_q} B_{\rm red}=F\ , 
\eeq
where ${\ovl \mu_q}$ is the average quark magnetic moment in the nonrelativistic constituent quark model, 
while the quark mass may be zero at this high density:  
\beq\label{4-3}
{\ovl \mu_q}=\frac{\mu_u+\mu_d}{2}\ , \qquad
\mu_u=\frac{\left(\frac{2}{3}e\right)\hbar}{2m_q}\ , \quad
\mu_d=\frac{\left(-\frac{1}{3}e\right)\hbar}{2m_q}\  
\eeq
with $\hbar$. 
Then, we obtain 
\beq\label{4-4}
{\ovl \mu_q}=\frac{1}{2}\cdot \frac{e\hbar}{2\cdot 3m_q}=\frac{1}{2}\frac{e\hbar}{2m_N}=1.576\times 10^{-17}\ {\rm GeV/T}\ ,
\eeq
where $3m_q=m_N$ is the nucleon mass. 
From Eq.(\ref{4-2}), the strength of the ``reduced" magnetic field is estimated. 
It may be expected that the ``reduced" magnetic field corresponds to the magnitude of ferromagnetization 
which occurs spontaneously in the high density quark matter. 
At high density, the order of the strength of ``reduced" magnetic field is about $10^{15}\sim 10^{16}$ T 
as is shown in Fig.\ref{fig:4-2}.

\section{Summary}

Quark ferromagnetization has been reinvestigated following the study\cite{Bohr1,Bohr2} 
performed by two of the present authors (J. da P. and C. P.) in the framework of 
the NJL type effective model of QCD. 
In this paper, the effective potential with respect to the condensate of quark spin alignment has been derived. 
Actually, it has been shown that, at high density, quark spin alignment occurs which leads to the spontaneous 
magnetization of the quark matter in the framework of the effective potential approach with the auxiliary field method 
based on the standard field theoretical technique. 
It is expected that non-vanishing condensate at high density leads to the effective field strength of spontaneous magnetization. 
In this paper, assuming the quark magnetic moment as that of the nonrelativistic constituent quark model, 
the strength of ``reduced" magnetic filed has been estimated where a rather large magnetization has been obtained. 

In this paper, having in mind the description of the interior of compact stars, the quark matter at zero temperature has been treated. 
Of course, it is interesting to investigate the system at finite temperature with the Fermi distribution function $n_+^{(\pm)}$. 
In this paper, we neglect the vacuum polarization by setting $n_-^{(\pm)}$ into $n_-^{(\pm)}-1$ simply to 
regularize the divergent integral. 
It may be interesting to study the effect of the vacuum polarization or of the Dirac sea for the spin 
polarization in quark matter at finite temperature.\cite{add1}
Further, at high density, the quark matter may reveal the color-superconducting feature. 
It is also interesting to examine the coexistence or competition between color-superconductivity and 
ferromagnetization\cite{add3,add4} in the framework developed in this paper.
These are future problems.

\section*{Acknowledgement} 

One of the authors (Y.T.) would like to express his sincere thanks to 
Professor\break
J. da Provid\^encia and Professor C. Provid\^encia, two of co-authors of this paper, 
for their warm hospitality during his visit to Coimbra in spring of 2012. 
Two of the authors (J.P. and C.P.) acknowledge valuable
discussions with H. Bohr and\break
P. K. Panda.
One of the authors (Y.T.) 
is partially supported by the Grants-in-Aid of the Scientific Research 
(No.23540311) from the Ministry of Education, Culture, Sports, Science and 
Technology in Japan.

\appendix
\section{Reinvestigation of the effective potential}

\begin{figure}[b]
\begin{center}
\includegraphics[height=4.5cm]{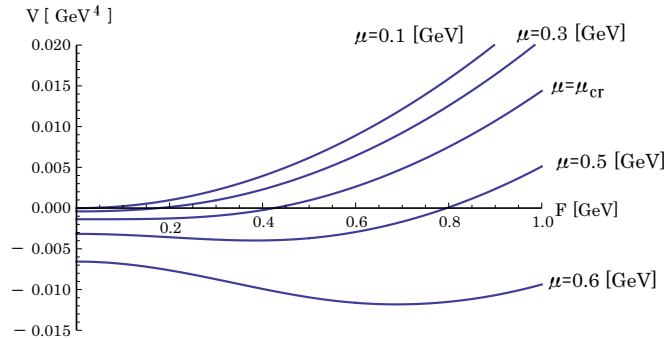}
\caption{
The effective potential is shown as a function of $F$.
From top, $\mu=0.1$ GeV, 0.3 GeV, $\mu_{\rm cr}$, 0.5 GeV and 0.6 GeV, respectively.
}
\label{fig:4-3}
\end{center}
\end{figure}

In this Appendix, we present one comment. 
In the definition of $V[F]$ in Eq.(\ref{3-8}) with the lower limit being 0 of integration with respect to $F$, 
this effective potential actually corresponds to 
$V[F]-V[F=0]$. If $V[F]$ itself is necessary as the effective potential, then we can express it as  
\beq\label{3-12}
V[F]=(V[F]-V[0])+V[0]\ . 
\eeq
Here, $V[F]-V[0]$ is finite as is seen from Eq.(\ref{3-8}). 
However, $V[0]=-\int d^4 p/(i(2\pi)^4)\cdot {\rm tr}\ln \gamma^\mu p_\mu$ is a part of the effective potential 
for the free massless quark gas and diverges due to the vacuum fluctuation. 
Therefore, in order to obtain the finite effective potential, it is necessary to replace the last term in Eq.(\ref{3-12}), 
$V[0]$, in an adequate way. 
One possibility is that $V[0]$ is replaced to the pressure $p_{\rm free}$ of a free massless 
quark gas:\cite{HTF}
\beq\label{3-13}
V[0]\rightarrow -p_{\rm free}=-N_cN_fT^4
\left[\frac{7\pi^2}{180}+\frac{1}{6}\left(\frac{\mu}{T}\right)^2+\frac{1}{12\pi^2}\left(\frac{\mu}{T}\right)^4
\right]\ . 
\eeq
Thus, we obtain a finite effective potential at $T=0$ as 
\beq\label{3-14}
V[F]
&=&\frac{F^2}{2G}-
\frac{N_c}{3\pi^2}\Biggl[
\frac{\sqrt{\mu^2-F^2}}{4}(3F^2\mu+2\mu^3)+F\mu^3{\rm arctan}\frac{F}{\sqrt{\mu^2-F^2}}
\nonumber\\
& &\qquad\qquad\qquad\qquad\qquad\qquad
-\frac{F^4}{4}
\ln\frac{\mu+\sqrt{\mu^2-F^2}}{F}\Biggl]
\eeq
with $N_f=2$. 
In this treatment, the effective potential for $F>\mu$ should be replaced by 
\beq\label{3-15}
V[F]=\frac{F^2}{2G}-\frac{N_c}{6\pi}\mu^3 F\ . \qquad ({\rm for}\ \ F>\mu)\ .
\eeq
In Fig.\ref{fig:4-3}, the effective potential defined by (\ref{3-14}) and (\ref{3-15}) is shown as a function of 
$F$. 
The gap equation is not changed. Thus, the results obtained in \S4  are retained.

\begin{figure}[t]
\begin{center}
\includegraphics[height=4.8cm]{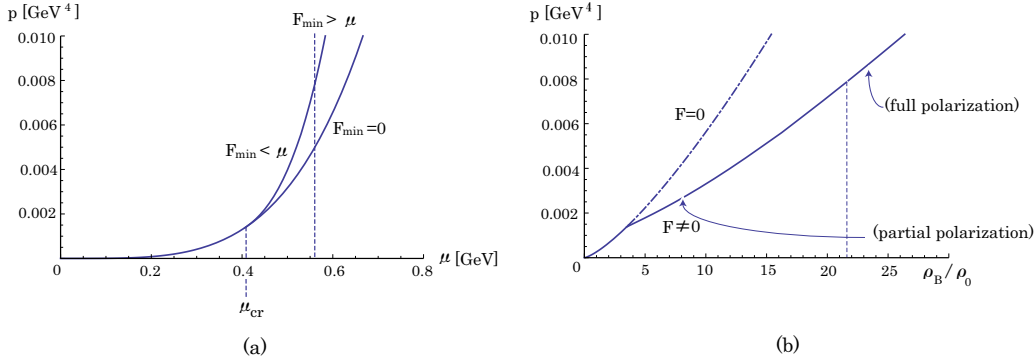}
\caption{
The pressure is shown as a function of (a) quark chemical potential $\mu$ and 
(b) baryon density divided by the normal nuclear density $\rho_0=0.17$ fm${}^{-3}$.
}
\label{fig:4-4}
\end{center}
\end{figure}

The effective potential itself may be regarded as the pressure $p$ in thermodynamics as is mentioned 
in Eq.(\ref{3-13}) already, namely, 
\beq\label{b1}
p=-V[F]\ . 
\eeq
In Fig.\ref{fig:4-4}, the pressure is shown as a function of (a) the quark chemical potential and 
(b) the baryon density divided by the normal nuclear density $\rho_0=0.17$ fm${}^{-3}$, respectively. 
As is seen from Fig.\ref{fig:4-4} (a), for quark chemical potential $\mu<\mu_{\rm cr}$, the branch with $F_{\rm min}=0$ 
is realized. 
From $\mu_{\rm cr}$ to $\mu=0.565$ GeV, partial magnetization with $F_{\rm min}<\mu$, in which 
the Fermi surface is close for the states with the single particle energy $\epsilon_p^{(+)}$ and 
these states do not contribute the magnetization, is realized. 
For large quark chemical potential, full magnetization with $F_{\rm min}>\mu$ is realized. 
As is seen from Fig.\ref{fig:4-4} (b), at low density, the normal phase with $F=0$ is realized. 
On the other hand, at high density, the quark ferromagnetic phase with $F\neq 0$ is realized. 
At the intermediate region between the low and the high density, 
partial magnetization is realized.

\section{The effect of vacuum polarization}

In this paper, we neglect the effect of the vacuum polarization because we replace 
$n_-^{(\pm)}$ into $n_-^{(\pm)}-1$, in which $n_-^{(\pm)}$ represents the negative energy contribution, 
as was mentioned in \S 3.
It may be important to investigate its influence for the ferromagnetization due to the axial vector interaction 
because the finite quark mass is essential in order to realize the ferromagnetization.\cite{add2} 
The existence of the finite quark mass is due to the chiral symmetry breaking which 
is generated by the quark-antiquark condensate. 
Thus, the effect of the vacuum polarization, which plays a crucial role for generating the chiral condensate, 
may be important.  
Namely, the Dirac sea plays an important role to realize the chiral condensate and the dynamical quark mass. 
However, the ferromagnetization occurs without quark mass in the case of tensor interaction under investigation in this paper.
Therefore, we simply subtract the negative energy contribution. 
It is well known that the role of the vacuum in models of the NJL type is very important 
and should not be ignored. 
The contribution $E_{\rm vac}=V_{\rm vac}[F,\Lambda]$ of the negative energy states, 
regularized by a cutoff $\Lambda$, can be easily obtained in our model. 
If we approximate this contribution by 
\beq\label{b0}
E_{\rm vac}\approx \frac{1}{2}F^2\frac{\delta ^2 V_{\rm vac}[F=0,\Lambda]}{\delta F^2}\ , 
\eeq
which seems reasonable, then the effect of the vacuum is the renormalization of the coupling constant $G$. 


In this Appendix, we focus on the transition point from normal to ferromagnetic phase. 
At zero temperature, $n_-^{(\pm)}=1$. 
Thus, we can calculate (\ref{2-20}) with $n_-^{(\pm)}=1$. 
As for the transition point, it is only necessary to calculate $\delta^2 V[F]/\delta F^2|_{F=0}$. 
We easily calculate it as 
\beq\label{b1}
\left. \frac{\delta^2 V[F]}{\delta F^2}\right|_{F=0}
=\frac{1}{G}-\frac{3\mu^2}{\pi^2}-\frac{\Lambda^2}{\pi^2}
\eeq
with $N_c=3$.
Thus, the result for the phase transition point is only obtained by replacing $1/G$ into a renormalized coupling 
$1/G_r=1/G-\Lambda^2/\pi^2$, 
namely, 
\beq\label{b2}
& &\left. \frac{\delta^2 V[F]}{\delta F^2}\right|_{F=0}
=\frac{1}{G_r}-\frac{3\mu^2}{\pi^2}\ , \nonumber\\
& &\frac{1}{G_r}=\frac{1}{G}-\frac{\Lambda^2}{\pi^2}\ . 
\eeq
Therefore, the qualitative behavior is not changed in the later discussion in the text, while the vacuum polarization is important 
for the spin polarization due to other interactions,\cite{add1} especially axial vector interaction.
If we adopt the standard value for the three momentum cutoff $\Lambda$, namely $\Lambda = 0.631$ GeV,\cite{HK} 
the critical baryon density$\rho_{\rm cr}$ and critical quark chemical potential $\mu_{\rm cr}$ are obtained numerically, 
which are summarized in Table {\ref{table:2}} instead of Table {\ref{table:1}} given in \S 4. 
It is seen that the contribution of the vacuum polarization to the critical point prefers the phase transition at rather low density. 
\begin{table}[b]
\caption{The critical baryon density $\rho_{\rm cr}$ and critical quark chemical potential $\mu_{\rm cr}$
by taking account of the vacuum polarization.}
\label{table:2}
\begin{center}
\begin{tabular}{c|c|c} \hline \hline
$G$ /GeV${}^2$ & $\rho_{cr}/\rho_0$ & $\mu_{\rm cr}$ /GeV \\ \hline
10 & 4.52 & 0.443  \\
11.028$(=2G_s)$ & 3.50 & 0.407 \\
15 & 1.32 & 0.294 \\
\hline
\end{tabular}
\end{center}
\end{table}
Here, the value $G=11.028$ corresponds to the strength obtained by the Fierz transformation 
as is discussed in \S 4. 
Further, around $F=0$, the effective potential itself is expressed as 
\beq\label{b3}
V[F]=\frac{F^2}{2G}+\cdots -\frac{\Lambda^2}{\pi^2}F^2=\frac{F^2}{2G_r}+\cdots\ .\qquad (|F|\ll 1)
\eeq
Then, the vacuum polarization effect can also be renormalized to the coupling in this case.



\end{document}